\documentclass[a4paper, 12pt, onecolumn]{article}

\title{\bf      Connected Operators for the
     Totally Asymmetric Exclusion Process
}
\author{      O. Golinelli, K. Mallick
\bigskip
\\ \ad        Service de Physique Th\'eorique, Cea Saclay, 91191 Gif, France
}

\date{\normalsize    6  April 2007}

\newcommand{\rop}[1]{\mathcal{O}\left(#1\right)} 

\newcommand{\binomial}[2]{{\scriptstyle #1 \choose \scriptstyle #2 }}

\newcommand  {\ad}{\normalsize\em}      
\begin{document}
\maketitle

\begin{abstract}
\normalsize

 We fully elucidate the structure of the hierarchy 
 of the connected operators that commute with the Markov
 matrix of the Totally Asymmetric Exclusion Process (TASEP).
  We prove  for the connected operators  a  combinatorial formula  that 
  was conjectured in a previous work.
 Our derivation is purely algebraic and relies 
 on the algebra generated by the local jump operators
 involved in the TASEP. 


\medskip \noindent 
Pacs numbers: 02.30.Ik, 02.50.-r, 75.10.Pq. 

\end{abstract}

 \section{Introduction}

     The Asymmetric Simple Exclusion Process (ASEP) is a lattice model
 of particles  with  hard core interactions. Due to its simplicity,
 the ASEP appears as a minimal model in many different contexts
  such as one-dimensional transport phenomena, molecular motors
 and traffic models. From a theoretical point of view,  this model 
 has become a paradigm   in the field of 
 non-equilibrium statistical mechanics;   many exact 
 results have been derived    using
 various methods, such as  continuous limits, Bethe Ansatz and 
 matrix Ansatz (for reviews, see e.g.,
 Spohn  1991,  Derrida 1998,  Sch\"utz 2001,  Golinelli and Mallick 2006). 

 In a recent work (Golinelli and Mallick 2007), 
 we applied   the algebraic  Bethe Ansatz technique
  to the Totally Asymmetric Exclusion Process (TASEP).
 This method allowed us to
   construct  a hierarchy of `generalized Hamiltonians'  that 
   contain the  Markov  matrix and commute with each other.
 Using the algebraic relations satisfied by the local jump
  operators,  we derived explicit  formulae
 for the transfer matrix and the generalized Hamiltonians, 
  generated from the transfer matrix.
  We showed  that the transfer matrix can be interpreted
 as the generator of a
  discrete time Markov process and we described the  actions 
 of  the   generalized  Hamiltonians.
 These actions  are non-local because they involve 
  non-connected  bonds of the lattice.
  However,  connected operators  
  are generated by taking
  the logarithm of the  transfer matrix.
   We  conjectured  for  the connected operators   a  combinatorial 
   formula  that was verified  for the   first ten connected operators
    by  using  a symbolic calculation program. 

 The aim of the present work is to present an analytical calculation 
 of the connected operators and  to prove the formula
 that was proposed in  (Golinelli and Mallick 2007).
  This paper is a sequel of our previous work,
 however, in section~\ref{sec:rappels}, we  briefly 
 review the main definitions and results  already obtained 
  so that this work can be read 
 in a fairly self-contained manner. 
  In section~\ref{sec:preuve}, we derive the general
 expression of the  connected operators.

 \section{Review of known results}
 \label{sec:rappels}
 We first recall the dynamical rules that define the TASEP 
 with $n$ particles 
 on a   periodic 1-d ring    with $L$ sites labelled $i=1,\dots,L$.
 The particles move   according to the following dynamics: during the time
interval $[t, t+dt]$,  a particle  on a site $i$   jumps with
probability $dt$ to the neighboring site $i+1$,  if this site  is empty.
This {\em exclusion
rule}  which  forbids to have more than one  particle  per  site,
 mimics a hard-core  interaction between particles.
 Because the particles can jump only in one direction this process
 is called totally asymmetric.
 The total  number $n$ of particles  is  conserved. The  TASEP  being 
  a continuous-time Markov process, its dynamics is entirely encoded 
in  a   $2^L \times 2^L$ Markov matrix $M$, that describes the evolution of
  the probability distribution of the system  at time $t$. 
 The   Markov matrix can be written as 
\begin{equation}
   M = \sum_{i=1}^L M_i \, , 
   \label{msm}
\end{equation}
where  the local jump operator
 $M_i$  affects only  the  sites $i$ and $i+1$
 and represents the contribution  to the dynamics
 of jumps from the site $i$ to $i+1$.
 
\subsection{The TASEP algebra}
\label{sec:TASEPalgebra}

 The local jump 
 operators satisfy a set of algebraic equations~:
\begin{eqnarray}
   M_i^2 &=& - M_i ,
   \label{mi2} \\    
   M_i \ M_{i+1} \ M_i = M_{i+1} \ M_i \ M_{i+1} &=& 0 ,
   \label{mmm=0} \\
   \left[M_i,M_j\right] &=& 0 \ \ \ \mathrm{if} \ |i-j| > 1 .
   \label{mimj}
\end{eqnarray} 
 These relations  can be obtained as a limiting form of the
Temperley-Lieb algebra. 
 On the ring we have periodic boundary conditions~: $M_{i+L} = M_{i}$.
The local jumps matrices  define an algebra. 
 Any  product of the  $M_i$'s will be called a {\it  word.}
 The {\it  length}  of a given  word is the minimal
   number of  operators $M_i$ required to write it.
    A word,   that can not be simplified further
   by using the algebraic rules above, will be called a reduced word. 

 Consider any  word  $W$   and  call  $\mathcal{I}(W)$ the set of indices  $i$
  of the  operators  $M_i$ that compose  it (indices are
  enumerated without repetitions). We remark that,  if  
  $W$ is not  annihilated  by application of rule~(\ref{mmm=0}), 
   the  simplification 
 rules~(\ref{mi2},  \ref{mimj})  do not alter  the set 
   $\mathcal{I}(W)$, {\it i.e.,} these  rules do not
 introduce any new index   or suppress any existing index
  in  $\mathcal{I}(W)$. This crucial property is not valid
  for the algebra associated
 with the  partially asymmetric exclusion process (see 
 Golinelli and Mallick 2006).

  Using the relation~(\ref{mi2}) we observe that for any $i$
 and any real number $\lambda \neq 1$ we have
\begin{equation}
  (1+\lambda M_i)^{-1} = (1+\alpha M_i)
  \ \ \ \mbox{ with } \ \ \ 
  \alpha = \frac{\lambda}{\lambda -1} \, .
  \label{alpha}
\end{equation}

\subsection{Simple words}

  A  simple word of length $k$
  is defined as  a word $M_{\sigma(1)} M_{\sigma(2)} \dots
M_{\sigma(k)}$, where $\sigma$ is a permutation on the set $\{1,2,
\dots, k\}$.  The  commutation  rule~(\ref{mimj}) implies that 
  only the relative position of $M_i$  with 
  respect to $M_{i \pm 1}$ matters.
 A  simple word  of length  $k$  can therefore be written
 as $W_k(s_2, s_3, \dots, s_k)$  where  the boolean
 variable  $s_j$ 
for  $ 2 \le j \le k$ is defined as follows~: 
 $s_j = 0$ if $M_j$ is on the left of  $M_{j-1}$
and   $s_j = 1$  if $M_j$ is on the right  of  $M_{j-1}$.
 Equivalently,  $W_k(s_2, s_3, \dots, s_k)$ 
 is uniquely defined
 by the recursion relation 
\begin{eqnarray}
   W_k(s_2, s_3, \dots, s_{k-1},1) &=& 
         W_{k-1}(s_2, s_3, \dots, s_{k-1}) \ M_k \ ,
 \label{def:simple1}\\
   W_k(s_2, s_3, \dots, s_{k-1},0)
    &=& M_k  \ W_{k-1}(s_2, s_3, \dots, s_{k-1}) \, .  \label{def:simple2}
\end{eqnarray}
   The set  of  the $2^{k-1}$simple words of length $k$ will be called
   $\mathcal{W}_k$.
 For a   simple word $W_k$,  we define  $u(W_k)$ to  be the number of {\em
inversions} in  $W_k$, {\it i.e.}, the number of times
that  $M_j$ is  on the left of $M_{j-1}$~: 
\begin{equation}
  u(W_k(s_2, s_3, \dots, s_k)) = \sum_{j=2}^{k} (1-s_j) \ .
 \label{def:u(W)}
\end{equation}

We remark that simple words are connected, they cannot be factorized
 in two (or more) commuting words.

\subsection{Ring-ordered product}

 Because of the periodic boundary
 conditions, products of local jump  operators must be ordered
 adequately.
 In the following we shall need to use  
 a  ring ordered product  $\rop{}$ which   acts on 
words   of the type   
\begin{equation}
 W = M_{i_1} M_{i_2} \dots M_{i_k} \, \hbox{   with  } \, 
  1 \le i_1 < i_2 < \dots < i_k \le L \, ,
 \end{equation}
   by   changing the  positions of matrices that appear in  $W$ 
   according to the following rules~:

     (i)  If $i_1 > 1$ or $i_k < L$, we define  $\rop{W} \ = W$. The  word
 $W$  is  well-ordered.

     (ii) If $i_1=1$ and $i_k=L$,
  we first   write  $W$ as a product of two blocks, $W = AB$,
  such that  $B = M_b
M_{b+1} \dots M_L$ is the maximal block of matrices with consecutive indices 
 that  contains  $M_L$,   and $A = M_1 M_{i_2} \dots M_{i_a}$, 
 with    $i_a < b-1$,  contains the  remaining terms. 
   We  then define
\begin{equation}
   \rop{W} =  \rop{AB} = BA 
        = M_b M_{b+1} \dots M_L M_1 M_{i_2} \dots M_{i_a} \, .
   \label{abba}
\end{equation}

  (iii) The previous  definition  makes sense  only for $k<L$.  Indeed, when 
  $k=L$, we have $W =M_1 M_2 \dots M_L$  and it is not possible 
  to split  $W$ in two different  blocks $A$ and $B$.
   For  this  special case,  we  define
\begin{equation}
  \rop{M_1 M_2 \dots M_L} \ = 
   | 1, 1, \dots, 1 \rangle \langle 1, 1, \dots, 1| \, , 
  \label{m1ml}
\end{equation}
 which  is  the projector on the   `full'   configuration with  all sites 
occupied.

  The ring-ordering  $\rop{}$  is extended by  linearity to  the vector space
 spanned by words of the type described above. 

\subsection{Transfer matrix and generalized Hamiltonians $H_k$}

  The algebraic Bethe Ansatz allows to construct a one parameter
  commuting family   of  transfer matrices,   $t(\lambda)$,
  that contains  the translation operator $T = t(1)$  and
 the Markov matrix $M = t'(0)$.  For $0 \le \lambda \le 1$, the operator  
  $t(\lambda)$  can be  interpreted 
  as a discrete time process with non-local jumps~:
 a hole located on the right of a cluster of $p$
particles can jump a distance $k$ in the backward direction, with
probability $\lambda^{k}(1-\lambda)$ for $1\le k < p $, and with
probability $\lambda^p$ for $k=p$.  The probability that this hole does
not jump at all is $1-\lambda$. This model  is equivalent
 to the  3-D anisotropic percolation model of Rajesh and Dhar 
 (1998) and  to a  2-D five-vertex model.
It is also an adaptation on a periodic lattice of the ASEP with a
backward-ordered sequential update (Rajewsky et al. 1996, Brankov et
al. 2004), and equivalently of an asymmetric fragmentation process
(R\'akos and Sch\"utz 2005).

 The  operator  $t(\lambda)$ 
 is a polynomial
  in  $\lambda$ of degree $L$ given by
\begin{equation}
     t(\lambda) =  1 + \sum_{k=1}^L \lambda^k H_k \ , 
   \label{thk}
\end{equation}
 where the generalized Hamiltonians  $H_k$ are {\it non-local operators}
 that act on the  configuration space.
 [We emphasize that   the notation  used here is different from that of 
  our previous work~:  $t(\lambda)$ was denoted by  $t_g(\lambda)$
 in (Golinelli and Mallick 2007).]
 
 We have $H_1 = M$ and
 more generally,  as shown  in (Golinelli and Mallick 2007),  
  $H_k$ is a homogeneous sum of words of length $k$ 
\begin{equation}
   H_k = \sum_{1 \le i_1 < i_2 < \dots < i_k \le L} 
          \rop{M_{i_1} M_{i_2} \dots  M_{i_k}}   \ ,  
   \label{hk}
\end{equation}
 where  $\rop{}$ represents the  ring ordered product
 that  embodies the periodicity and the translation-invariance
constraints.

 For a system of size L with $N$ particles
  only $H_1, H_2, \ldots, H_{N}$ have a non-trivial action.
 Because we are interested only in the case 
  $N \le L-1$ (the full system as no dynamics)
  there are at most $L-1$ operators $H_k$
 that have a non-trivial action.

 \section{The connected operators $F_k$}
  \label{sec:preuve}

 \subsection{Definition}

The generalized Hamiltonians  $H_k$  and the transfer
 matrix   $t(\lambda)$  have  non-local actions
 and couple  particles with arbitrary distances
 between them. Besides  $H_k$ is a highly non-extensive
 quantity as it involves generically a number of terms
 of  order $L^k$.  As usual, the  local connected and
  extensive operators are obtained by taking
 the logarithm of the transfer matrix. 
  For   $k\ge 1$,   the 
connected Hamiltonians  $F_k$  are defined as
\begin{equation}
  \ln t(\lambda) = \sum_{k=1}^{\infty} \frac{\lambda^k}{k} F_k \ .
  \label{lntg}
\end{equation}
 Taking the derivative of this equation
 with respect to $\lambda$ and recalling that
 $t(\lambda)$ commutes with  $t'(\lambda)$, we obtain 
\begin{equation}
  \sum_{k=1}^{\infty} \lambda^k F_k 
  = \lambda \ t(\lambda)^{-1}\ t'(\lambda)
  \label{t-1t'} \, . 
\end{equation}
  Expanding  $t(\lambda)^{-1}$  with respect to $\lambda$, 
 this formula allows to calculate  $F_k$ as a  polynomial function of 
 $H_1,\ldots,H_k$.  For example  
  $F_1 = H_1$, $F_2 = 2H_2 - H_1^2$, etc... (see Golinelli and Mallick 2007).
   By using~(\ref{hk}), we observe that   $F_k$  is  {\it a priori} a
 linear combination of 
  products of $k$ local operators  $M_i$.  However this expression 
  can   be simplified  by using the 
 algebraic rules~(\ref{mi2}, \ref{mmm=0}, \ref{mimj}) and  {\it in fine},
  $F_k$ will be a linear combination of reduced  words of length $j \le k$.
 
  Because of the ring-ordered product
 that appears in the expression~(\ref{hk}) of the $H_k$'s, 
 it is difficult to derive an expression  of   $F_k$ in terms
 of the local jump operators.
 An exact formula  for the  $F_k$ with $k \le 10$ was obtained 
 in (Golinelli and Mallick 2007) by using a computer program
 and a general expression was conjectured for all $k$.
   In the following, the  conjectured formula  
    is derived and  proved rigorously.

 \subsection{Elimination of the ring-ordered product}

 The expression $\sum \lambda^{k} F_k$
 can  be written as a linear combination of  reduced words $W$.
  We know
 from  formula~(\ref{hk}) that  at most 
 $L-1$ operators $H_k$  are independent  in  a system of size L, 
 we shall therefore  calculate  $F_k$ only for $k \le {L-1}$. Thus,  
 we  need  to  consider   reduced words of length  $j \le L-1$. 
 Let   $W$ be  such a word, and $ \mathcal{I}(W)$ 
 be the set of  indices of the operators $M_i$ that compose  $W$; 
  our  aim  is 
 to find the expression  of  $W$  and to 
  calculate its prefactor from  equation~(\ref{t-1t'}).
  Because the rules~(\ref{mi2},  \ref{mimj})
   do not suppress or add any new index, the following
  property is true~:  if a word $W'$ appearing in  
 $\lambda \ t(\lambda)^{-1}\ t'(\lambda)$  is such that
  $  \mathcal{I}(W') \neq  \mathcal{I}(W)$ then  even
  after simplification,  $W'$ will remain different from 
   $W$.  
 Therefore, the prefactor  of  $W$ in  $\sum \lambda^{k} F_k$ 
  is the same as the prefactor  of  $W$ in 
\begin{equation}
  \lambda \ t_{\mathcal{I}}(\lambda)^{-1}\ t'_{\mathcal{I}}(\lambda)
 \,\,\, \hbox{ where }  t_{\mathcal{I}}(\lambda) 
       = \rop{\prod_{i \in \mathcal{I}} ( 1 + \lambda M_i)} \, 
 \,\,\,\, \hbox{ with } \mathcal{I}(W) \subset  \mathcal{I} \, .  
  \label{titi}
\end{equation}

  Because  $F_k$  commutes with the translation operator $T$,  
  then for any $r=1, \dots, L-1$,   the prefactor  of   $W =
  M_{i_1} M_{i_2} \dots M_{i_j}$ is the same as the  prefactor  of
  $ T^r M T^{-r}  =  M_{r+i_1} M_{r+i_2} \dots M_{r+i_j}$.  Furthermore, any
  word  $W$ of size $k \le L-1$ is equivalent, by 
  a translation,  to a word that contains  $M_1$ and not $M_L$~:
  indeed, there exists at least one index  $i_0$  such that 
 $i_0 \notin \mathcal{I}(W)$ and  $ (i_0+1) \in \mathcal{I}(W)$ 
 and it is thus sufficient to translate $W$  by $r = L - i_0$.  

  In conclusion,  
 it is enough to  study in    expression~(\ref{t-1t'}), 
  the reduced  words  $W$  with  set of indices included in 
\begin{equation}
  \mathcal{I}^{*}  = \{1, 2, \dots, L-1 \} \, . 
\end{equation}
 Because the index  $L$ does not appear in  $\mathcal{I}^{*}$, 
  the ring-ordered product has a trivial action in equation~(\ref{titi})
 and we have 
\begin{equation}
   t_{\mathcal{I}^{*}}(\lambda) =
    (1 + \lambda M_1) (1 + \lambda M_2) \dots (1 + \lambda M_{L-1}) \, .
 \label{eq:noROP}
\end{equation}
   We have thus been able to eliminate the ring-ordered product.

 \subsection{Explicit formula for the connected operators}

   In equation~(\ref{eq:noROP}), differentiating 
 $t_{\mathcal{I}^{*}}(\lambda)$
 with respect to $\lambda$,  we have
\begin{equation}
   t'_{\mathcal{I}^{*}}(\lambda) =
      \sum_{i=1}^{L-1}
        (1 + \lambda M_1) \dots (1 + \lambda M_{i-1}) M_i 
        (1 + \lambda M_{i+1}) \dots (1 + \lambda M_{L-1}) \, .
\end{equation}
 Using equation~(\ref{alpha}) we obtain
\begin{equation}
  t_{\mathcal{I}^{*}}(\lambda)^{-1} = 
  (1 + \alpha M_{L-1}) (1 + \alpha M_{L-2}) \dots (1 + \alpha M_1) \, ,
 \,\, \hbox{with } \,\,  \alpha = \frac{\lambda}{\lambda-1} \, .
\end{equation}
 Noticing that 
 $\lambda(1 + \alpha M_i) M_i = - \alpha M_i$, we deduce 
\begin{eqnarray}
  &&\lambda \
  t_{\mathcal{I}^{*}}(\lambda)^{-1}\ t'_{\mathcal{I}^{*}}(\lambda) = 
   \\ 
   &&-\alpha  \sum_{i=1}^{L-1}
   (1 + \alpha M_{L-1}) \dots (1 + \alpha M_{i+1})
   M_i
   (1 + \lambda M_{i+1}) \dots (1 + \lambda M_{L-1}) \, .\nonumber
\end{eqnarray}
The $i$th  term in this sum   contains words with
 indices between $i$ and $L-1$.  Because we  are looking for the 
   words that contain the operator  $M_1$, we must 
  consider only the first term in this sum, which we note by $Q$
\begin{equation}
  Q = - \alpha (1 + \alpha M_{L-1}) \dots (1 + \alpha M_2)
      M_1
      (1 + \lambda M_2) \dots (1 + \lambda M_{L-1}) \, . 
\label{defQ}
\end{equation}
 In the appendix, we show that
\begin{equation}
Q = R_1 + R_2 + \dots + R_{L-1} \, ,
\label{eqQR}
\end{equation}
 where $R_i$  is  defined    by the  recursion~:
\begin{eqnarray}
  R_1 &=& -\alpha M_1 \ ,
\\
  R_i &=& \lambda R_{i-1} M_i + \alpha M_i R_{i-1} 
 \,\,\,\, \hbox{ for } i \ge 2 \, .
\label{def:ri}
\end{eqnarray}
 To summarize,   all  the words in 
  $\sum_{k=1}^{\infty} \lambda^k F_k$ that contain 
  $M_1$ and not  $M_L$   are given by 
 $Q = R_1 + R_2 + \dots + R_{L-1}$. 
 From the recursion relation~(\ref{def:ri})
 we deduce that  $R_i$ is a  linear combination of  the 
 $2^{i-1}$ simple  words  $W_i(s_2, s_3, \dots,s_i)$
 defined in section~\ref{sec:TASEPalgebra}. Furthermore,
 we observe from~(\ref{def:ri}) that   
  a factor $\lambda$  appears if $s_i = 1$
 and a factor  $\alpha = \lambda/(\lambda -1)$   appears if $s_i = 0$.
 Therefore, the coefficient  $f(W)$  of  $W = W_i(s_2, s_3, \dots,s_i)$
  in  $Q$ is given by 
\begin{equation}
  f(W) = (-1)^u \frac{ \lambda^i}{(1 -\lambda)^{u+1}}
=   (-1)^u \sum_{j=0}^{\infty} \binomial{u+j}{j} \lambda^{i+j}
 \label{eq:fW}
\end{equation}
 where  $i$ is the length of $W$ and
  $u = u(W)$  is its  inversion number,  defined in
 equation~(\ref{def:u(W)}). We have thus shown that
\begin{equation}
  Q = \sum_{i=1}^{L-1} \sum_{W \in \mathcal{W}_i}  f(W)\  W
 = \sum_{i=1}^{L-1} \sum_{W \in \mathcal{W}_i}  W \sum_{j=0}^{\infty}
  (-1)^{u(W)}  \binomial{u(W)+j}{j} \lambda^{i+j}  \, ,
\label{eq:formuleQ}
\end{equation} 
where $\mathcal{W}_i$ is the set of simple words of length $i$.

 Finally, we recall that the coefficient  
  in $\sum_{k=1}^{\infty} \lambda^k F_k$ 
  of a  reduced
  word $W$ that contains $M_1$ and not $M_L$
  is the same as its
 coefficient in $Q$.   Extracting  the term of order $\lambda^k$
 in equation~(\ref{eq:formuleQ}) we deduce that any word $W$  in
  $F_k$  that contains $M_1$ and not $M_L$  is a  simple word 
  of length  $i \le k $ and its prefactor is given by 
 $ (-1)^{u(W)}  \binomial{u(W)+k-i}{k-i}$. 

 The full expression of $F_k$ is obtained by applying the
 translation operator to the expression~(\ref{eq:formuleQ});
  indeed any word in  $F_k$  can be uniquely  obtained
  by translating a simple word in  $F_k$  that contains $M_1$ and not $M_L$. 
We conclude that for   $k < L$, 
\begin{equation}
  F_k = \mathcal{T} \sum_{i=1}^k \sum_{W \in \mathcal{W}_i}
            (-1)^{u(W)} \binomial{k-i+u(W)}{k-i} W \ ,
  \label{fk}
\end{equation}
 where $\mathcal{T}$ is the translation-symmetrizator 
 that  acts on any operator $A$ as follows~:
$  \mathcal{T} A  = \sum_{i=0}^{L-1}  T^{i} \ A \  T^{-i}\,. $ 
  The presence of   $\mathcal{T}$   in equation~(\ref{fk})
  insures that  $F_k$ 
  is invariant by translation on the periodic system of
  size $L$. All simple words being 
  connected, we  finally  remark that  
 formula~(\ref{fk}) implies  that $F_k$ is a connected operator.

 \section{Conclusion}
  
   By using the algebraic properties of the 
 TASEP algebra~(\ref{mi2}-\ref{mimj}), we have derived an
 exact combinatorial
 expression for the family of connected operators that
 commute with the Markov matrix. This calculation allows
 to fully elucidate the hierarchical structure
  obtained 
 from the Algebraic Bethe Ansatz. 
 It would be  of a great interest to extend  our result 
 to the partially asymmetric exclusion process (PASEP),
 in which a particle can make forward jumps with probability $p$ and
  backward jumps with probability $q$. 
 In particular, we recall
  that the symmetric exclusion
 process is equivalent to the Heisenberg spin chain~: in this case
  the connected  operators have  been calculated  only 
 for the lowest orders (Fabricius et al., 1990). 
 This is a  challenging  and difficult problem. In our derivation 
 we used a fundamental  property of  the  TASEP algebra~: 
 the  rules~(\ref{mi2}-\ref{mimj})
 when applied to a word $W$ either cancel  $W$  or conserve
  the set of indices $\mathcal{I}(W)$. 
 The algebra associated with  PASEP violates this crucial 
  property  because  there  we have 
  $M_i \ M_{i+1} \ M_i =  pq \ M_i .$
 Therefore the method followed here does not
 have a straightforward extension to the PASEP case.


\appendix

\section*{Appendix: Proof of equation~(\ref{eqQR})}

 Let us define  the  following series
\begin{eqnarray}
  Q_1 &=& -\alpha M_1 \ ,
\\
  Q_i &=& (1 + \alpha M_i) Q_{i-1} (1 + \lambda M_i)
 \,\,\,\, \hbox{ for } i \ge 2 \, .
  \label{qi}
\end{eqnarray}
We remark that  $Q$  defined in equation~(\ref{defQ})
 is given by  $Q = Q_{L-1}$.
 Let us consider   $R_i$ defined   by the  recursion~(\ref{def:ri}).
The indices that appear in the words
 of   $Q_i$ and  $R_i$ 
    belong to $\{1, 2,\dots, i\}$.  Therefore, we have
\begin{equation}
  [R_j, M_i] = 0 
  \ \ \ \mbox{ for } \ \ \
  j \le i-2 \, ,
  \label{rm}
\end{equation}
because the operators  $M_1, M_2, \dots, M_j$ that
 compose  $R_j$  commute  with  $M_i$. 
 From  equations~(\ref{rm}) and  (\ref{alpha}), we obtain
\begin{equation}
  (1 + \alpha M_i) R_j (1 + \lambda M_i) = R_j \,\,\,\,
 \hbox{ for }   j \le i-2 \, .
\label{eqRj}
\end{equation}
 Furthermore,  from~(\ref{def:ri}), we obtain 
\begin{equation}
  M_i R_{i-1} M_i 
  = \lambda M_i R_{i-2} M_{i-1} M_i 
   + \alpha M_i M_{i-1} R_{i-2} M_i  \, .
\end{equation}
 Because  $M_i$ commutes  with $R_{i-2}$,  we can use
 the relation  $M_i M_{i-1} M_i = 0$ to deduce that
\begin{equation}
  M_i R_{i-1} M_i = 0 \, .
  \label{mrm}
\end{equation}
 
 Using  equation~(\ref{mrm}), we find
\begin{equation}
  (1 + \alpha M_i) R_{i-1} (1 + \lambda M_i) = 
  R_{i-1} + \lambda R_{i-1} M_i + \alpha M_i R_{i-1}
  = R_{i-1} + R_i \, .
\label{eqRim1}
\end{equation}

  From equations~(\ref{eqRj}) and~(\ref{eqRim1}), we  prove that 
 the (unique) solution of the recursion relation~(\ref{qi})
 is given by equation~(\ref{eqQR}), 
  $Q_i = R_1 + R_2 + \dots + R_i .$

\section*{References}

\begin{itemize}

\item
  Brankov J. G., Priezzhev V. B. and Shelest R. V., 2004,
  {\em Generalized determinant solution of the discrete-time totally
  asymmetric exclusion process and zero-range process}, 
  Phys. Rev. E {\bf 69} 066136.

\item Derrida B., 1998, {\em An exactly soluble non-equilibrium
    system: the asymmetric simple exclusion process}, Phys. Rep.  {\bf
    301}  65.

\item  Fabricius~K., M\"utter~K.-H. and Grosse~H., 1990,
 {\em  Hidden symmetries in the one-dimensional antiferromagnetic
 Heisenberg model}, Phys. Rev. B {\bf 42}  4656.

\item
Golinelli O. and Mallick K.,  2006,
{\em  The asymmetric simple exclusion process: an integrable model
 for non-equilibrium statistical mechanics,}
  J. Phys. A: Math. Gen. {\bf 39}  12679.

\item
Golinelli O. and Mallick K.,  2007,
{\em  Family of Commuting Operators for the
     Totally Asymmetric Exclusion Process,} 
  Submitted to  J. Phys. A: Math. Theor.,
  cond-mat/0612351.

\item Rajesh~R. and  Dhar~D., 1998,   {\em An exactly solvable anisotropic
  directed percolation model in three dimensions},  
    Phys. Rev. Lett. {\bf 81}  1646.  

\item
  Rajewsky N., Schadschneider A. and Schreckenberg M., 1996,
  {\em The asymmetric exclusion model with sequential update},
  J. Phys. A: Math. Gen. {\bf 29} L305.

\item
  R\'akos A. and Sch\"utz G. M., 2005, 
  {\em Current distribution and random matrix ensembles for an integrable
  asymmetric fragmentation process},
  J. Stat. Phys. {\bf 118} 511.

\item   Sch\"utz~G.~M., 2001,
{\em Exactly solvable models for many-body systems far from equilibrium}
 in {\em Phase Transitions and Critical Phenomena,} vol. 19, C. Domb and
 J.~L.~Lebowitz Ed., Academic Press, San Diego.

\item  Spohn~H., 1991,
{\em Large scale dynamics of interacting particles},
 Springer, New-York.

\end{itemize}

\end{document}